\RequirePackage{ifpdf}
\ifpdf % We are running pdfTeX in pdf mode
\documentclass[pdftex]{sigma}
\else
\documentclass{sigma}
\fi

\begin{document}

\renewcommand{\PaperNumber}{031}

\FirstPageHeading

\renewcommand{\thefootnote}{$\star$}

\ShortArticleName{Recent Applications of the Theory of Lie Systems in Ermakov Systems}

\ArticleName{Recent Applications of the Theory of Lie Systems\\ in Ermakov Systems\footnote{This paper is a contribution to the Proceedings
of the Seventh International Conference ``Symmetry in Nonlinear
Mathematical Physics'' (June 24--30, 2007, Kyiv, Ukraine). The
full collection is available at
\href{http://www.emis.de/journals/SIGMA/symmetry2007.html}{http://www.emis.de/journals/SIGMA/symmetry2007.html}}}

\Author{Jos\'e F. CARI\~NENA, Javier  DE LUCAS and Manuel F. RA\~NADA}

\AuthorNameForHeading{J.F. Cari\~nena, J. de Lucas and M.F. Ra\~nada}

\Address{Department of Theoretical Physics, University of Zaragoza, 50.009
  Zaragoza, Spain}
\Email{\href{mailto:jfc@unizar.es}{jfc@unizar.es}, \href{mailto:dlucas@unizar.es}{dlucas@unizar.es}, \href{mailto:mfran@unizar.eso}{mfran@unizar.es}}

\ArticleDates{Received November 02, 2007, in f\/inal form February
04, 2008; Published online March 12, 2008}

\Abstract{We review some recent results of the theory of Lie systems
 in order to apply such results to study Ermakov
  systems.
 The fundamental properties of Ermakov systems, i.e.\ their superposition rules,
 the Lewis--Ermakov invariants, etc., are found from this new
 perspective. We
also obtain new results, such as a new superposition rule for the Pinney
equation in
 terms of three solutions of a related Riccati equation. }


\Classification{34A26; 34A05}

\section{Introduction}

Nonlinear equations have been of an increasing  interest in Physics during the last
thirty years and one of the simplest examples is the today called
Milne--Pinney equation
\begin{gather}\label{MP}
\ddot x=-\omega^2(t)x+\frac k{x^3},
\end{gather}
where $k$ a is real constant with values depending on the f\/ield in which the
equation is to be applied.
Such equation was introduced by  an Ukrainian mathematician of the nineteenth
century, V.P.~Ermakov,
as a way of looking for a f\/irst  integral for the  time-dependent
 harmonic oscillator \cite{Er80}. He was using  some of Lie's ideas for dealing
 with ordinary dif\/ferential equations with tools of  classical
geometry. Lie had obtained a
characterisation of non-autonomous systems of f\/irst-order dif\/ferential
equations
\begin{gather}
\frac{dx^i}{dt }= Y^i(t,x),\qquad  i=1,\ldots,n,\label{nonasys}
\end{gather}
admitting a superposition rule \cite{LS}. Such a problem has being receiving
these last years very much attention  because of its very important
applications in physics and mathematics, see~\cite{NI1,NI2,Win83,CGM,CarRamGra,CM01,And80,HarWinAnd83,
OlmRodWin8687,KD,LKNA05,OlmRodWin87,CarRamdos,CarRamcinc,CMN98}.
This approach has recently been revisited from a more geometric perspective in~\cite{CGM07}
where the r\^ole of the superposition function is played by an appropriate
connection.
This new approach allows us to consider a superposition of
 solutions of a
given system in order to obtain solutions of another system as it was done in
 \cite{CLR07}. Our aim in this paper is to
show how such a superposition rule may be understood from a geometric viewpoint
in some interesting cases, as the Milne--Pinney equation (\ref{MP})
 \cite{{Mil30},{P50}}, the Ermakov system and its
 generalisations.

Recall that Ermakov systems are systems of second-order dif\/ferential equations composed by
the  Milne--Pinney dif\/ferential equation (\ref{MP})
together with  the corresponding time-dependent harmonic oscillator.
They have received lot of attention since its introduction  in  Ermakov's paper~\cite{Er80},
where many of its important properties have been exhibited. For
instance, he found the time-dependent invariant called Ermakov--Lewis
invariant. In a short paper Pinney~\cite{P50}  was able to prove the
fundamental result
 that the general solution of the Milne--Pinney equation can be described in
 terms of an arbitrary  pair of independent particular solutions of the
corresponding time-dependent harmonic oscillator and two constants.

  Ermakov systems have also  been broadly studied in Physics since its
  introduction  until now. They  appear in the
 study of the Bose--Einstein condensates and cosmological models \cite{HL02,
   Li04,Ha02}, and in the solution of time-dependent harmonic or anharmonic
 oscillators  \cite{FM03, Ga84, RR79a, DL84, Le67}. Much  work has also
 been
 devoted to use Hamiltonian or Lagrangian structures in the  study of such a
 system, see e.g.~\cite{RR80,{Cerveropla156}}, and many
 generalisations or new insights from the mathematical point of view can be found in
 \cite{WS78, GAL93, ARRO90, SaCa82, RR80b, GL94a, At99, RSB96, PGL91}.

Our aim in this paper is to review the mathematical insight developed in
\cite{CGM}
in order to study the properties of Ermakov systems. With  this objective,
we review in Section~\ref{SR}
the
concept of superposition rule, a map describing the general solution of a
certain kind of system of dif\/ferential equations called  Lie systems,  in terms of a f\/inite
set of particular solutions of it. The concept  is generalised  in Section~\ref{sec3} to express
 the general solution of a system of second-order
dif\/ferential equations (hereafter shortened as SODE) in terms of arbitrary particular solutions of a new Lie
system
and some constants.  We prove that the theory developed by Lie
can easily be adapted to deal with many examples of such SODE systems, and
in this way the
superposition rules which are generally used for f\/irst-order dif\/ferential
equations can  also  be  used for dealing with  Ermakov systems.
We f\/ind room in this way
for implicit nonlinear superposition rules in the
terminology of \cite{RR80b,WS81}. The Ermakov--Lewis invariants \cite{Le67}
appear in a natural way as functions def\/ining the foliation associated to the
superposition rule.  Some simple examples are used to illustrate the theory.
Section~\ref{SR2} is devoted to describe the  possibility of a~superposition of solutions
of a~system to give a~solution of another related one, and we show how Pinney's
results \cite{P50} f\/it in this framework, i.e.\ we recover the known
superposition
 rule in which the solutions for the Pinney equation are expressed in terms of
two solutions of the corresponding  $t$-dependent harmonic oscillator and as a
straightforward application of our method we show that the general solution of
the Pinney equation can also  be written as a superposition of three particular solutions of the Riccati equation.

\section[Systems of differential equations admitting a superposition
  rule]{Systems of dif\/ferential equations\\ admitting a superposition
  rule}\label{SR}

A superposition rule for solutions of (\ref{nonasys}) is  determined by a function
$\Phi:{\mathbb{R}}^{n(m+1)}\to {\mathbb{R}}^n$,
\begin{gather*}
x=\Phi(x_{(1)}, \ldots,x_{(m)};k_1,\ldots,k_n),\qquad x_{(a)}\in\mathbb{R}^n,\qquad a=1,\ldots,m,%\label{superpf}
\end{gather*}
such that the general solution can be written, for
suf\/f\/iciently small $t$, as
\begin{gather*}
x(t)=\Phi(x_{(1)}(t), \ldots,x_{(m)}(t);k_1,\ldots,k_n) ,%\label{superpft}
\end{gather*}
where $\{x_{(a)}(t)\mid a=1,\ldots,m\}$ is a fundamental set of
particular solutions of the system (\ref{nonasys}) and
$k=(k_1,\ldots,k_n)$ is a set of $n$  arbitrary constants, whose values f\/ix each particular solution. A~superposition rule provides us a
method for f\/inding the general solution of (\ref{nonasys})
in terms of a~set of particular solutions.

According to  the
Implicit Function Theorem, the function $\Phi:\mathbb{R}^{n(m+1)}\to\mathbb{R}^n$
can be, at least locally
around generic points, inverted, so we can write
\begin{gather*}
k=\Psi(x_{(0)}, \ldots,x_{(m)}) %\label{defPsi}
\end{gather*}
for a certain function $\Psi:{\mathbb{R}}^{n(m+1)}\to {\mathbb{R}}^n$.
Hereafter in order to handle a shorter notation we started writing $x_{(0)}$
instead of $x$. The level sets of the superposition function $\Psi$
def\/ine a~foliation that is invariant
under permutations of the $(m+1)$ variables. The functions~$\Phi$ and~$\Psi$
are related~by:
\begin{gather}
k=\Psi(\Phi(x_{(1)}, \ldots,x_{(m)};k_1,\ldots,k_n),x_{(1)}, \ldots,x_{(m)}) .\label{implth}
\end{gather}
The fundamental property of the superposition function $\Psi$ is that as
\begin{gather*}
k=\Psi(x_{(0)}(t),x_{(1)}(t), \ldots,x_{(m)}(t)),%\label{kconstancy}
\end{gather*}
the function  $\Psi(x_{(0)}, \ldots,x_{(m)})$ is constant along any
$(m+1)$-tuple of solutions of the system (\ref{nonasys}). This implies that if the
 `diagonal prolongation' $\widetilde
Y(t,x_{(0)}, \ldots, x_{(m)})$ of the $t$-dependent  vector f\/ield
$Y(t,x)=Y^i(t,x)\partial/\partial x^i$ is def\/ined by
\[
\widetilde Y(t,x_{(0)}, \ldots,x_{(m)})=\sum_{a=0}^mY_a(t,x_{(a)}),\qquad
t\in {\mathbb{R}},
\]
where
\begin{gather*}%\label{Ya}
Y_{a}(t,x_{(a)})=\sum_{i=1}^n Y^i(t,x_{(a)}) \frac{\partial}{\partial x^i_{(a)}},
\end{gather*}
then $\widetilde
Y(t,x_{(0)}, \ldots, x_{(m)})$ is a $t$-dependent  vector f\/ield on
${\mathbb{R}}^{n(m+1)}$
 tangent to the level sets of $\Psi$, i.e.\ the  components $\Psi^i$ are
constants of motion. The level sets of $\Psi$
corresponding to
regular values def\/ine a $n$-codimensional foliation $\mathcal{F}$ on an open
dense subset $U\subset {\mathbb{R}}^{n(m+1)}$
and the family $\{\widetilde Y(t,\cdot),\,t\in {\mathbb{R}}\}$ of vector
f\/ields in ${\mathbb{R}}^{n(m+1)}$   consists of vector f\/ields tangent to
the leaves of this foliation.

Remark that, as pointed out in \cite{CGM07}, for any
$k=(k_1,\ldots,k_n)\in{\mathbb{R}}^n$ and each  $(x_{(1)},\ldots,x_{(m)}) \in {\mathbb{R}}^{nm}$   there is one
 point $(x_{(0)},x_{(1)},\ldots,x_{(m)})$ on the level
set $\mathcal{F}_k$ of the foliation $\mathcal{F}$,
namely, such that
$(\Phi(x_{(1)},\ldots,x_{(m)};k),x_{(1)},\ldots,x_{(m)})\in
\mathcal{F}_k $ (cf.~(\ref{implth})); then, the projection onto the
last $m$ factors
\[
{\rm pr}:(x_{(0)},x_{(1)},\ldots,x_{(m)})\in {\mathbb{R}}^{n(m+1)}\mapsto
(x_{(1)},\ldots,x_{(m)})\in {\mathbb{R}}^{nm}
\]
induces dif\/feomorphisms on the leaves $\mathcal{F}_k$ of $\mathcal{F}$.
Such a foliation gives us
the superposition principle without referring to the function
$\Psi$: if we f\/ix the point $x_{(0)}(0)$
 and $m$ solutions
$x_{(1)}(t),\ldots,x_{(m)}(t)$, then $x_{(0)}(t)$ is the unique
point in ${\mathbb{R}}^n$ such that
$(x_{(0)}(t),x_{(1)}(t),\ldots,x_{(m)}(t))$ belongs to the same
leaf of $\mathcal{F}$ as
$(x_{(0)}(0),x_{(1)}(0),\ldots,x_{(m)}(0))$. This means that  only
$\mathcal{F }$ really matters for the superposition rule.

Lie's main result \cite{LS} can be expressed as follows:
\medskip

\noindent
{\bf Lie's Theorem.}  {\it The system \eqref{nonasys} on a differentiable manifold $N$
admits a superposition rule if and only if the $t$-dependent vector field
$Y(t,x)$ can be locally written in the form
\begin{gather}
Y(t,x)=\sum_{\alpha=1}^r b_\alpha(t)\, X_\alpha(x) ,\label{Liesys}
\end{gather}
where the vector fields $X_\alpha$, with $\alpha=1,\dots,
r$, close on a $r$-dimensional real Lie algebra, i.e.\ there
exist $r^3$ real numbers $c_{\alpha\beta}\,^\gamma $ such that{\samepage
\begin{gather}
[X_\alpha,X_\beta]= \sum_{\gamma=1}^r c_{\alpha\beta}\,^\gamma
X_\gamma ,\qquad \forall\, \alpha,\beta=1,\ldots,r .\label{Liealg}
\end{gather}
In this case we say that \eqref{nonasys} is a Lie system.}}

The number $m$ of solutions involved in the superposition rule for the
Lie system def\/ined by~(\ref{Liesys})
with generic $b_\alpha(t)$ is the
minimal $k$ such that the diagonal prolongations
of $X_1,\dots,X_r$ to $N^k$, $\widetilde
X_1,\dots,\widetilde X_r$, are linearly independent at (generically)
each point: for $k=m$ the only real numbers solution of the linear system
\begin{gather*}
\sum_{\alpha=1}^r c_\alpha\, \widetilde X_\alpha(x_{(1)}, \ldots, x_{(k)})=0 ,
\end{gather*}
at a generic point $(x_{(1)},\dots,x_{(k)})$ is the trivial
solution $c_\alpha=0$, with $\alpha=1,\ldots,r$,  and there
are nontrivial solutions for $k<m$. Then, the superposition function $\Psi$ is
made up of $t$-independent constants of motion for the prolonged $t$-dependent
vector f\/ield $\widetilde Y$. We shall call them f\/irst-integrals.

A possible  generalisation consists on considering foliations  of codimension
dif\/ferent from
$n$, or even choosing dif\/ferent Lie systems with the same associated Lie
algebra
for def\/ining the prolonged vector f\/ield. These facts will be illustrated with
several examples in the following sections.

\section{SODE Lie systems}\label{sec3}

A system of second-order dif\/ferential equations
\[
\frac{d^2x^i}{dt^2}=F^i\left(t,x,\frac{dx}{dt}\right),\qquad i=1,\ldots,n
\]
can be studied through the system of f\/irst-order dif\/ferential equations
\begin{gather*}
\dot x^i = v^i,\\
\dot v^i = F^i(t,x,v),
\qquad i=1,\ldots,n,
\end{gather*}
with an associated $t$-dependent vector f\/ield
\[X=v^i\frac{\partial}{\partial x^i}+F^i(t,x,v)\frac{\partial }{\partial  v^i} .\]

We call SODE Lie systems those SODE for which $X$ is a Lie system, i.e.\ $X$ can be
written as a~linear combination with $t$-dependent coef\/f\/icients of vector f\/ields
closing a f\/inite-dimensional real Lie algebra.  Even if we are focusing here
our attention on second-order dif\/ferential equations, the approach can be
straightforwardly applied for $r$-order dif\/ferential equations.
 Some particular examples of second-order Lie systems are given in next subsections.

\subsection[Milne-Pinney equation]{Milne--Pinney equation}

The   Milne--Pinney  equation  (\ref{MP}) is a  second-order nonlinear dif\/ferential  equation
\cite{{Mil30},{P50}}
that  describes the time-evolution of an
 isotonic oscillator \cite{Cal69,Pe90} (also called pseudo-oscillator), i.e.\ a
harmonic  oscillator with inverse
 quadratic potential \cite{WS78}. This nonlinear oscillator shares with the harmonic one
 the property of
having a period independent of the energy \cite{ChaVes05}, i.e.\ it is
a~isochronous system,
 and its quantum version has an equispaced spectrum \cite{ACMP07,CPR07}.

Note that  if $x(t)$ is solution of (\ref{MP}) then  $x'(t)=-x(t)$ is
solution too and  we can restrict ourselves to  consider solutions in the
half-line  $\mathbb{R}_+=\{x\in \mathbb{R}\mid  x>0\}$. Thus,  we are
interested in the solutions of (\ref{MP}) that are curves in $\mathbb{R}_+$.

 We  can relate the Milne--Pinney equation with a system of f\/irst-order
dif\/ferential equations def\/ined in ${\rm T}\mathbb{R}_+$ by introducing a new
auxiliary variable
 $v\equiv \dot x$. Such system is given by
\begin{gather}
\dot x = v,\nonumber\\
\dot v = -\omega^2(t)x+\dfrac{k}{{x^3}},
 \label{MPsys}
\end{gather}
where $x\in \mathbb{R}_+$ and $(x,v)\in {\rm T}_x\mathbb{R}_+$. By the sake of simplicity we assume from now on that $k>0$ but similar results can be obtained for $k<0$. Now, the associated $t$-dependent vector f\/ield over~${\rm T}\mathbb{R}_+$ is
\[X=v\frac{\partial}{\partial x}+\left(-\omega^2(t)x+\frac k{x^3}\right)\frac{\partial}{\partial v} .
\]
This is a Lie system because $X$ can be written as
\[X=L_2-\omega^2(t)L_1 ,
\]
where the vector f\/ields $L_1$ and $L_2$ are given by
\[
L_1=x\frac {\partial}{\partial v},\qquad L_2=
v\frac{\partial}{\partial x}+\frac k {x^3}\frac{\partial}{\partial v},
\]
which are such that
\[
[L_1,L_2]=2L_3,\qquad [L_3,L_2]=-L_2,\qquad [L_3,L_1]=L_1
\]
with
\[
 L_3=\frac 1 2 \left(x\frac{\partial}{\partial x}-v\frac{\partial}{\partial v}\right) ,
\]
i.e.\ they span a 3-dimensional real  Lie algebra $\mathfrak{g}$
  isomorphic  to $\mathfrak{sl}(2,\mathbb{R})$.  Another dif\/ferent relation of
  the Lie algebra $\mathfrak{sl}(2,\mathbb{R})$  with this  system has been pointed out but in the context of
  Lie symmetries \cite{LK04,  KY02}. On the contrary here $\mathfrak{sl}(2,\mathbb{R})$
is the Lie algebra of vector f\/ields of the Lie system and the search for the
general solution of (\ref{MPsys}) can
be reduced to f\/inding the solution starting from the neutral element of the
equation in the group $SL(2,\mathbb{R})$ given by
$\dot g\, g^{-1}=\omega^2(t) {\rm a_1}- {\rm a_2}$ (see
\cite{Win83,CGM}), where $\{{\rm a_1},{\rm a_2},{\rm a_3}\}$
is the following
 basis of $\mathfrak{sl}(2,\mathbb{R})$:
\begin{gather}
{\rm a}_1=\left(\begin{array}{cc}
0&0\\
-1&0
\end{array}\right),\qquad
{\rm a}_2=\left(\begin{array}{cc}
0&-1\\
0&0
\end{array}\right),\qquad
{\rm a}_3=\frac 12\left(\begin{array}{cc}
-1&0\\
0&1
\end{array}\right) ,\label{basissl2}
\end{gather}
which satisfy the same commutation relations as the vector f\/ields $\{L_j\mid j=1,2,3\}$. Actually, it is possible to show that $L_j$ is the fundamental vector f\/ield corresponding to ${\rm a}_j$ with respect to the  action $\Phi: (A,(x,v))\in SL(2,\mathbb{R})\times {\rm T}\mathbb{R}_+\rightarrow (\bar x,\bar v)\in {\rm T}\mathbb{R}_+$ given by:
\begin{gather*}
\bar x =\sqrt{\dfrac{k+\left[(\beta v+\alpha x)(\delta
      v+\gamma x)+ k({\delta\beta}/{x^2})\right]^2}{(\delta
    v+\gamma x)^2+k({\delta}/{x})^2}},\\
\bar v =\kappa \sqrt{\left(\delta v+\gamma
  x\right)^2+\dfrac{k\delta^2}{x^2}\left(1-\dfrac{x^2}{\delta^2\bar
    x^2}\right)} \qquad {\rm if}\quad A\equiv\left(\begin{array}{cc}
\alpha & \beta\\ \gamma &\delta \end{array}\right),
\end{gather*}
where $\kappa$ is $\pm 1$ or $0$ depending on the initial point $(x,v)$ and the
element of the group $SL(2,\mathbb{R})$ that acts on it. In order to obtain an
explicit expression for $\kappa$ in terms of $A$ and $(x,v)$ we can use the following decomposition for any element of the group $SL(2,\mathbb{R})$
\[
A=\exp(-\alpha_2{\rm a}_2)\exp(-\alpha_1{\rm a}_1)\exp(-\alpha_3{\rm a}_3)=
\left(\begin{array}{cc}
      1&\alpha_2\\
0&1
      \end{array}\right)
\left(\begin{array}{cc}
      1&0\\
\alpha_1&1
      \end{array}\right)
\left(\begin{array}{cc}
      e^{\alpha_3/2}&0\\
0&e^{-\alpha_3/2}
      \end{array}\right),
\]
from where we obtain that $\alpha_1=\gamma\delta$ and
$\alpha_2=\beta/\delta$. As we know that $\Phi(\exp(-\alpha_3{\rm a}_3),(x,v))$
is the integral curve of the vector f\/ield $L_3$ starting from  the point $(x, v)$
parametrised by $\alpha_3$, it is straightforward to check that
\[(x_1,v_1)\equiv\Phi(\exp(-\alpha_3{\rm a}_3),(x,v))=(\exp(\alpha_3/2)x,
\exp(-\alpha_3/2)v),\]
and in a similar way \[(x_2,v_2)\equiv\Phi(\exp(-\alpha_1{\rm a}_1),(x_1,v_1))=(x_1,\alpha_1 x_1+v_1).\]

Finally, we want to obtain $(\bar x,\bar v)=\Phi(\exp(-\alpha_2{\rm
  a}_2),(x_2,v_2))$, and taking into account that the integral curves of $L_2$
satisfy that
\begin{gather}\label{intcur}
\frac{x^3dv}{k}=\frac{dx}{v}=d\alpha_2,
\end{gather}
it turns out that when
$k>0$ we have $\bar v^2+k/\bar x^2=v_2^2+k/x_2^2\equiv\lambda$ with
$\lambda>0$. Thus,
 using  this fact and (\ref{intcur}) we obtain
\[
\frac{k^{1/2}dv}{(\lambda-v^2)^{3/2}}=d\alpha_2,
\]
and integrating $v$ between $v_2$ and $\bar v$,
\begin{gather*}
\frac{\bar v}{(\lambda-\bar v^2)^{1/2}}=\alpha_2\frac{\lambda}{k^{1/2}}+\frac{v_2}{(\lambda-v_2^2)^{1/2}}
=\frac{1}{k^{1/2}}\left(\alpha_2\lambda+v_2|x_2|\right).
\end{gather*}
As $\kappa={\rm sign}[\bar v]$, we see that  $\kappa$ is given by
\[
\kappa={\rm sign}[\alpha_2\lambda+v_2|x_2|]={\rm sign}\left[\frac\beta\delta(x\gamma+v\delta)^2+\frac{k\delta\beta}{x^2}+\frac{|x|}{\delta}(v\delta+x\gamma)\right].
\]

 Non-trivial f\/irst-integrals independent of $\omega(t)$ for
the equation  (\ref{MP}) do not exist, i.e.\ there is not any $t$-independent
 constant of motion $I:U\subset{\rm T}\mathbb{R}_+\rightarrow\mathbb{R}$ such
 that $XI=0$ for any function~$\omega(t)$. This is equivalent to $dI(L_j)=0$
 for
  $L_j$ with $j=1,2,3$. Thus,
 the integrals of motion we are looking  for verify that $dI$ vanishes on
 the involutive
 distribution
$\mathcal{V}(x)\simeq \langle L_1(x),L_2(x),L_3(x)\rangle$ generated by the
fundamental vector f\/ields $L_j$. In almost any point we have that
$\mathcal{V}(x)={\rm T}_x{\rm T}\mathbb{R}_+$, and  as $dI=0$ on dense subsets
in ${\rm T}\mathbb{R}_+$, the only possibility is $dI=0$, therefore $I$ is a
constant, and
the integral of motion is trivial.

\subsection{Generalised Ermakov system}

Consider a  possible  generalisations of the Ermakov system given by
\begin{gather}
\ddot{x}=-\omega^2(t)x+\frac{1}{x^3}f(y/x),\nonumber\\
\ddot{y}=-\omega^2(t)y+\frac{1}{y^3}g(y/x).\label{SOrder}
\end{gather}
This system of dif\/ferential equations has been broadly studied in
 \cite{RR79a,RR80,SaCa82,GL94a,WS81,RR79b,R81}.
 In this section we analyse how  can  we make use of the theory of
Lie systems in order to obtain the known f\/irst-integral of motion for this
system of dif\/ferential equations.

Note that  if $(x(t),y(t))$ is a solution, then   $(-x(t),-y(t))$ is also a solution
 for (\ref{SOrder}), and therefore we can focus our attention on the solutions
of generalised Ermakov system which are curves in the manifold
$\mathbb{R}_+\times \mathbb{R} $.

The generalised Ermakov system (\ref{SOrder}) can be written as a system
 of f\/irst-order dif\/ferential
equations with
a double number of
degrees of freedom by introducing two new variables~$v_x$ and~$v_y$,
\begin{gather}
\dot x =v_x,\nonumber\\
\dot v_x =-\omega^2(t)x+\frac 1 {x^3} f( y/x),\nonumber\\
\dot y =v_y,\nonumber\\
\dot v_y =-\omega^2(t)y+\frac 1 {y^3} g(y/x),\label{FOrder}
\end{gather}
and  its solutions can be studied from  the integral curves of the
$t$-dependent vector f\/ield $X$ in ${\rm T}(\mathbb{R}_+\times \mathbb{R})$ given by
\[X=v_x\,\frac{\partial}{\partial x}+v_y\,\frac{\partial}{\partial y}+\left(-\omega^2(t)x+\frac 1 {x^3} f( y/
  x)\right)\frac{\partial}{\partial v_x}+\left(-\omega^2(t)y+\frac 1 {y^3} g(y/
    x)\right)\frac{\partial}{\partial v_y} ,
\]
which can be written as a linear combination
\[X=N_2-\omega^2(t)  N_1,
\]
where $N_1$ and $N_2$ are the vector f\/ields
\[
N_1=x\frac{\partial }{\partial v_x}+y\frac{\partial }{\partial v_y},\qquad N_2=v_x\frac{\partial}{\partial x}+v_y\frac{\partial}{\partial y}+
\frac{1}{x^3}f(y/x)\frac{\partial}{\partial v_x}+
\frac{1}{y^3}g( y/x)\frac{\partial}{\partial v_y} .
\]

Note that these vector f\/ields generate a 3-dimensional real Lie algebra
with a third generator
\[N_3=\frac 12\left(x\frac{\partial}{\partial
    x}+y\frac{\partial}{\partial
    y}-v_x\frac{\partial}{\partial v_x}-v_y\frac{\partial}{\partial v_y}\right) .\]

In fact, as
\[
[N_1,N_2]=2N_3, \qquad [N_3,N_1]=N_1, \qquad  [N_2,N_3]=N_2 ,
\]
they generate a Lie algebra isomorphic to   $\mathfrak{sl}(2,\mathbb{R})$. Therefore the
generalised Ermakov system is a SODE Lie system with associated Lie algebra $\mathfrak{sl}(2,\mathbb{R})$.

The  integrable distribution  associated to the Lie system (\ref{FOrder}) is
of
dimension  three, while
the manifold ${\rm T}(\mathbb{R}_+\times \mathbb{R})$   is
 4-dimensional, and then  there exists  a f\/irst-integral of motion, $F:\mathbb{R}^4\rightarrow \mathbb{R}$,
for any $\omega^2(t)$. This f\/irst-integral $F$  satisf\/ies
$N_iF=0$ for $i=1, 2, 3$, but as $[N_1,N_2]=2N_3$ it is enough to impose
$N_1F=N_2F=0$. Then, if $N_1F=0$, by the method of characteristics we can conclude that
 there exists a function $\bar F:\mathbb{R}^3\rightarrow \mathbb{R}$ such that
$F(x,y,v_x,v_y) =\bar F(x,y,\xi)$ with $\xi=xv_y-yv_x)$.
 The
condition   $N_2F=0$ reads now
\[
v_x\frac{\partial\bar F}{\partial x}+v_y\frac{\partial\bar F}{\partial
  y}+\left(-\frac{y}{x^3}f({y}/{x})+
\frac{x}{y^3}g({y}/{x})\right)\frac{\partial \bar F}{\partial \xi}=0 .
\]

We can therefore consider the associated system of
 the characteristics:
\[
\frac{dx}{v_x}=\frac{dy}{v_y}=\frac{d\xi}{-\frac{y}{x^3}f({y}/{x})+\frac{x}{y^3}g({y}/{x})},
\]
and using that
\[
\frac{-y\,dx+x\,dy}{\xi}=\frac{dx}{v_x}=\frac{dy}{v_y} ,
\]
we arrive to
\[
\frac{-y\,dx+x\,dy}{\xi}=\frac{d\xi}{-\frac{y}{x^3}f(\frac{y}{x})+\frac{x}{y^3}g(\frac{y}{x})},
\]
i.e.
\[
-\frac{y^2d\left(\frac{x}{y}\right)}{\xi}=
\frac{d\xi}{-\frac{y}{x^3}f(\frac{y}{x})+\frac{x}{y^3}g(\frac{y}{x})}
\]
and integrating we obtain the following f\/irst-integral
\[
\frac 12 \xi^2+\int^{x/y}\left[-\frac 1{\zeta^3}\, f\left(\frac 1\zeta\right)+
  \zeta\,g\left(\frac 1\zeta
\right)\right] d\zeta=C  .
\]
 This
f\/irst-integral allows us  to determine, by means of quadratures,  a solution of
a subsystem
 in terms of  a solution of the other equation.

\subsection{The  harmonic oscillator with time-dependent frequency}

Our aim in this section is to exhibit  with a simple example that
we can def\/ine new Lie systems of dif\/ferential equations  admitting
 f\/irst-integrals of motion which do not depend on the time-dependent
 coef\/f\/icients by putting together
some copies of the same Lie system in the way indicated   in Section
\ref{SR}. These integrals of motion allow us to obtain relations between
 dif\/ferent solutions of the initial Lie systems and can be used to construct a
 superposition rule if  enough
 copies  have been   added.

 The equation of motion of a 1-dimensional harmonic oscillator with time-dependent
 frequency is $\ddot x= -\omega^2(t) x$, which is a
second-order dif\/ferential equation whose solutions
are curves in $\mathbb{R}$. We can alternatively consider the
following system of f\/irst-order dif\/ferential equations
\begin{gather}\dot x = v,\nonumber \\
\dot v = -\omega^2(t)
    x,\label{1dimho}
\end{gather}
whose solutions are the integral curves in ${\rm T}\mathbb{R}$ of the
 $t$-dependent vector f\/ield
\[X=v\frac{\partial}{\partial x} -\omega^2(t) x  \frac{\partial}{\partial v} .
\]
Now  $X$ is a linear combination $X=X_2- \omega^2(t)X_1$, where $X_1$ and $X_2$
 are vector f\/ields  in ${\rm T}\mathbb{R}$:
\[X_1= x\frac{\partial}{\partial v} ,\qquad X_2=v  \frac{\partial}{\partial x} ,
\]
such that
\begin{gather}\label{CR}
[X_1,X_2]=2 X_3 , \qquad [X_3,X_1]= X_1 ,\qquad [X_2,X_3]=X_2 ,
\end{gather}
where $X_3$ is the vector f\/ield  given by
\[
X_3=\frac 12 \left(x\frac{\partial}{\partial x}-v\frac{\partial}{\partial v}\right)  .
\]

As a consequence of  Lie's Theorem we see that $X$ def\/ines a Lie system with associated  Lie algebra
$\mathfrak{sl}(2,\mathbb{R})$. Actually, the vector f\/ields $\{X_\alpha\mid \alpha=1,2,3\}$ are
 fundamental vector f\/ields corresponding to the
usual linear action  of $SL(2,\mathbb{R})$ on $\mathbb{R}^2\simeq {\rm T}\mathbb{R}$ given by
\begin{gather*}
\bar x=\alpha x+\beta v,\qquad
\bar v = \gamma x+\delta v
 \qquad {\rm if} \qquad A\equiv\left(\begin{array}{cc}
\alpha & \beta\\ \gamma &\delta
\end{array}\right).
\end{gather*}

Remark that  the elements of the basis of $\mathfrak{sl}(2,\mathbb{R})$
given by (\ref{basissl2})  close the same commutation relations as
 the vector f\/ields $\{X_\alpha\mid \alpha=1,2,3\}$ which turn out to be the
 associated fundamental vector f\/ields for the elements ${\rm a}_\alpha$.

Non-trivial f\/irst-integrals independent of
$\omega(t)$ for the system (\ref{1dimho}) do not exist, because the fundamental vector f\/ields generate the involutive
distribution $\mathcal{V}\simeq \langle X_1,X_2,X_3\rangle$ whose values in each
point of ${\rm T}\mathbb{R}$ generate the corresponding tangent space
in almost any point and then as~$dI$ vanishes on
 $\mathcal{V}$,  therefore $dI$ vanishes in a generic point. Thus
 $I$ is a
trivial f\/irst integral of motion.

As it was said before,  non-trivial integrals of motion are needed  to
develop the procedures of integration indicated in   \cite{CGM}, i.e.\ we should
add  copies of the same dif\/ferential equation
until the linear subspace of the involutive distribution of the total system
in a generic point has a~dimension smaller
than the dimension of the manifold. In this way there is room for a non-constant
function~$I$
such that~$dI$ vanishes on  the subspace of the distribution associated with
the
extended  system of dif\/ferential equations.

Thus, we f\/irst consider a system with two copies of the same dif\/ferential equation,
\begin{gather}
\ddot x_1 =  -\omega^2(t) x_1,\nonumber\\
\ddot x_2 =  -\omega^2(t)   x_2 ,\label{2dimho}
\end{gather}
which corresponds to a 2-dimensional  isotropic
 harmonic oscillator with a time-dependent frequency and  is
associated with the following system of f\/irst-order dif\/ferential equations
\begin{gather}
\dot x_1 = v_1,\nonumber\\
 \dot v_1 = -\omega^2(t) x_1,\nonumber\\
\dot x_2 = v_2, \nonumber\\
 \dot v_2 =  -\omega^2(t) x_2,
\label{2dimhob}
\end{gather}
whose solutions are the integral curves of the
 $t$-dependent  vector f\/ield
\[X=v_1\frac{\partial}{\partial x_1}+v_2\frac{\partial}{\partial x_2}
 -\omega^2(t) x_1\, \frac{\partial}{\partial v_1}
 -\omega^2(t) x_2\, \frac{\partial}{\partial v_2}  ,
\]
which is a linear combination, $X=X_2- \omega^2(t) X_1$, with $X_1$ and $X_2$
being the vector f\/ields
\[X_1= x_1\frac{\partial}{\partial v_1}+ x_2\frac{\partial}{\partial v_2}
 ,\qquad X_2=v_1 \frac{\partial}{\partial x_1}+v_2 \frac{\partial}{\partial x_2} ,
\]
which are such that
\begin{gather}\label{CR2}
[X_1,X_2]=2  X_3 , \qquad [X_3,X_1]= X_1  ,\qquad [X_2,X_3]=X_2 ,
\end{gather}
where the vector f\/ield $X_3$ is def\/ined by
\[X_3=\frac 12 \left(x_1\frac{\partial}{ \partial
    x_1}+x_2\frac{\partial}{\partial x_2}
-v_1\frac{\partial}{\partial v_1}
-v_2
\frac{\partial}{\partial v_2}
\right)  .
\]

The system of dif\/ferential equations (\ref{2dimhob}) is therefore a Lie system,
i.e.\
the $t$-dependent vector f\/ield can be written  as a linear combination with
time-dependent coef\/f\/icients of vector f\/ields~$X_\alpha$ closing a f\/inite
dimensional Lie algebra. Such vector f\/ields are  diagonal prolongations of the fundamental
vector f\/ields of the initial system of dif\/ferential equation and satisfy  the same commutation relations~(\ref{CR2})
 as (\ref{CR}). The subspace of the distribution associated for this Lie system
 has a rank lower or equal to the dimension of the Lie algebra. The Lie algebra
 associated
with the $t$-dependent vector f\/ield $X$ is also
 $\mathfrak{sl}(2,\mathbb{R})$.

Note that the system  (\ref{1dimho}) did not admit
  non-trivial f\/irst-integral
 of motion indepen\-dent of~$\omega(t)$, but the system  (\ref{2dimho}) does admit
 a f\/irst-integral because the subspace of the distribution generated by the fundamental
 vector f\/ields  has rank three in almost any point  and the dimension of the
 total
 manifold is four. Thus, there exists a function  $F$ such that $dF$ is
 vanishes on~the distribution generated by the $X_\alpha$'s. The function
$F(x_1,x_2,v_1,v_2)$ is such that
$X_1F=0$ if\/f there exists a function      $\bar F(x_1,x_2,\xi)$ with
$\xi=x_1v_2-x_2v_1$,
such that $F(x_1,x_2,v_1,v_2)=\bar F(x_1,x_2,\xi)$, and~then the second
condition
$X_2F=0$ implies that $\bar F$ can only depend on $\xi$, i.e.\
$F(x_1,x_2,v_1,v_2)=\hat F(\xi)$. From the commutation relation
 $2  X_3=[X_1,X_2]$, we see that the conditions $X_1F=X_2F=0$ imply $X_3F=0$,
 and
therefore any f\/irst integral is a function of
$F(x_1,x_2,v_1,v_2)=x_1v_2-x_2v_1$, which physically corresponds to the angular
momentum. This f\/irst integral can be seen from the  mathematical viewpoint
  as a partial superposition rule~\cite{CGM}. Actually, if $x_1(t)$ is a
solution of the f\/irst equation, then we obtain for each real number $k$
 the f\/irst-order dif\/ferential
equation for the variable $x_2$
\[x_1(t)  \frac{dx_2}{dt} =k+\dot x_1(t)x_2 ,
\]
from where $x_2$ can be found to be given by
\begin{gather*}
x_2(t)=k' x_1(t)+k  x_1(t)\int^t \frac{d\zeta}{x_1^2(\zeta)},%\label{redunasol}
\end{gather*}
where $k'$ is a new integration constant.

In order to look for a superposition rule we should
consider a system of  some copies of  (\ref{1dimho}) with  at least
as
many integrals of motion as the dimension of the initial manifold. Then
it may be possible to obtain the variables of the initial manifold explicitly in
terms of the other variables. Following the development of the Section~\ref{SR}
a way to obtain the number of
necessary particular solutions  to obtain a superposition rule is to consider a set of copies of the initial system and check out if the prolongations of the vector f\/ields $X_1$, $X_2$ and $X_3$ are linearly independent in a~generic point.

In the case of two copies of the $t$-dependent harmonic oscillator it is possible to show that in a generic point if $\lambda_1 \tilde X_1+\lambda_2  \tilde X_2+\lambda_3  \tilde X_3$ vanishes then
$\lambda_1=\lambda_2=\lambda_3=0$, therefore $m= 2$ and  consequently
 there is a superposition rule
involving two particular solutions. In order to obtain the superposition
rule we need to consider three copies of the $t$-dependent harmonic oscillator
and
 study the system of f\/irst-order dif\/ferential equations
\begin{gather*}\dot x_1 = v_1,\nonumber\\
 \dot v_1 =  -\omega^2(t) x_1,\nonumber
\dot x_2 = v_2,\nonumber\\
 \dot v_2 =  -\omega^2(t) x_2,\nonumber\\
 \dot x = v, \nonumber\\
  \dot v =  -\omega^2(t) x,
%\label{3dimhob}
\end{gather*}
whose solutions are the integral curves of the
 $t$-dependent vector f\/ield
\[X=v_1\frac{\partial}{\partial x_1} +v_2\frac{\partial}{\partial x_2}
+v\frac{\partial}{\partial x}
-\omega^2(t) x_1  \frac{\partial}{\partial v_1}
 -\omega^2(t) x_2  \frac{\partial}{\partial v_2}
-\omega^2(t) x  \frac{\partial}{\partial v}  ,
\]
which is a linear combination, $X=X_2- \omega^2(t) X_1$, with $X_1$ and $X_2$
being the vector f\/ields
\[X_1= x_1\frac{\partial}{\partial v_1}+ x_2\frac{\partial}{\partial v_2}+ x\frac{\partial}{\partial v}
\,,\qquad X_2=v_1 \frac{\partial}{\partial x_1}+v_2 \frac{\partial}{\partial x_2}+v \frac{\partial}{\partial x} ,
\]
which are such that
\[[X_1,X_2]=2  X_3 , \qquad [X_3,X_1]= X_1  ,\qquad [X_2,X_3]=X_2 ,
\]
where the vector f\/ield $X_3$ is def\/ined by
\[X_3=\frac 12 \left(x_1\frac{\partial}{ \partial x_1}
+x_2\frac{\partial}{\partial x_2}+x\frac{\partial}{\partial x}
-v_1\frac{\partial}{\partial v_1}
-v_2 \frac{\partial}{\partial v_2}
-v
\frac{\partial}{\partial v}
\right)  .
\]

 The f\/irst-integrals $F$ are as in the last case
the  solutions of $X_1F=X_2F=0$, because $2 X_3=[X_1,X_2]$. The condition $X_1F=0$ says that there exists a
function $\bar F:{\mathbb{R}}^5\to {\mathbb{R}}^2$ such that $F(x_1,x_2,x,v_1,v_2,v)=\bar F (x_1,x_2,x,\xi_1
,\xi_2)$ with $\xi_1(x_1,x_2,x,v_1,v_2,v)=xv_1-x_1v$ and
$\xi_2(x_1,x_2,x,v_1,v_2,v)=xv_2-x_2v$, and when written in terms of $\bar F$  the condition
$X_2F=0$ implies that $\bar F$ is an arbitrary function of
 $\xi_1$ and $\xi_2$  (of course, $\xi=x_1v_2-x_2v_1$ is
also a f\/irst-integral). They produce a superposition
rule, because from
\begin{gather*}
 xv_2-x_2v= k_1,\qquad
  x_1v-v_1x = k_2
\end{gather*}
 we obtain the expected superposition rule for two solutions:
\[x=c_1  x_1+c_2  x_2,\qquad v=c_1 v_1+c_2 v_2,\qquad
c_i=\frac{k_1}{k}, \qquad  k=x_1v_2-x_2v_1 .
\]

\section{Construction of mixed superposition rules}\label{SR2}

In the preceding sections  we have considered SODE Lie systems together with  some copies
of the same system. In this section we show a way to deal with
sets of maybe dif\/ferent systems of dif\/ferential equations sharing  the same associated
Lie algebra as Lie systems. These
sets satisfy certain conditions to be explained
later on and  can be used to
 obtain mixed superposition rules.

Suppose that a Lie system is given
\[
\frac{dx}{dt}=\sum_{\alpha=1}^rb_\alpha(t)X_\alpha(x),
\]
where the $X_\alpha$ with $\alpha=1,\ldots,r$, close a f\/inite-dimensional Lie
algebra as given by (\ref{Liealg}).
We want~to obtain a mixed superposition rule for this Lie system, i.e.\ a function
$\Phi: N^1\times\cdots\times N^m\times\mathbb{R}^n\rightarrow N$ in such a way
that any
solution
integral curve $x(t)$ of $X(t,\cdot)$ be given by
\[
x(t)=\Phi(x_{(1)}(t),\dots,x_{m}(t),k_1,\dots,k_n),
\]
where $x_{(1)}(t),\dots, x_{(m)}(t)$, are integral curves of the $t$-dependent vector f\/ields on the correspon\-ding manifolds $N^{j}$ with $j=1,\ldots, m$ and given by
\begin{gather*}
X^{(1)}(t,\cdot) =\sum_{\alpha=1}^rb_\alpha(t)X^{1}_\alpha(\cdot),\\
\cdots \cdots \cdots \cdots\cdots\cdots \cdots\cdots  \cdots  \\
X^{(m)}(t, \cdot) =\sum_{\alpha=1}^rb_\alpha(t)X^{m}_\alpha(\cdot)
\end{gather*}
and in such a way that $X^{j}_\alpha$ with $j=1,\ldots,m$, and $\alpha=1,\ldots, r$, close the same commutation
relations as the vector f\/ields $X_\alpha$, i.e.
\[
[X^j_\alpha,X^j_\beta]=c_{\alpha\beta}\,^{\gamma}X^j_\gamma,\qquad j=1,\ldots,m, \qquad{\rm and}\qquad \alpha,\beta=1,\ldots r.
\]

The search for  a set of vector f\/ields closing  a given Lie algebra in a
certain manifold amounts to look for dif\/ferent actions of the corresponding
Lie group.

Then, over $\tilde N$ we obtain that
\begin{gather*}
[X_\alpha,X^j_\beta]=0,\qquad [X^j_\alpha,X^k_\beta]=0,\qquad j,k=1,\ldots,m,\quad
{\rm with} \quad j\ne k,\! \quad{\rm and}\quad \alpha,\beta=1,\ldots r,
\end{gather*}
and because of  these commutation relations we can def\/ine in $\tilde N$ the
vector f\/ields
\[
\tilde Y_\alpha=X_\alpha+\sum_{i=1}^{m}X^{i}_\alpha
\]
which satisfy  the same commutation relations as the vector f\/ields $X_\alpha$ in (\ref{Liealg}).
Thus, the  system of dif\/ferential equations determining the integral curves of the $t$-dependent vector f\/ield
\[
\tilde Y(t,\cdot)=\sum_{\alpha=1}^{r}b_\alpha(t){\tilde Y}_\alpha(\cdot)
\]
is a Lie system with the same associated Lie algebra as the initial system of
dif\/ferential equations. Nevertheless, the dimension of the manifold is this
time  larger than the dimension of $N$ because we have put together several
manifolds.
The distribution $\tilde{\mathcal{V}}$ associated to this Lie system is given by
$
\tilde{\mathcal{V}}(\tilde x)=\langle  {\tilde Y}_1(\tilde x),\ldots,{\tilde Y}_r(\tilde x)\rangle
$
and is involutive. The dimension of  the subspace in a point is lower or equal
to $r$
and therefore when the dimension of
the manifold $\tilde N$ is larger than $r$ there are integrals of motion for any
value  of the time-dependent coef\/f\/icients which may be used to obtain
 superposition rules.

In the next section we  give some examples of the application of  this
procedure
 to the Ermakov system in order to both recover  previously known properties
 and
to f\/ind also new results.

\subsection{Ermakov system}\label{ES}

Consider the system of ordinary f\/irst-order dif\/ferential equations \cite{DL84,PGL91}
\begin{gather*}
\dot x = v_x,\\
\dot v_x = -\omega^2(t)x,\\
\dot y = v_y,\\
\dot v_y = -\omega^2(t)y+\dfrac {1}{{y^3}}%\label{Ermak} \nonumber
\end{gather*}
made up by  a 1-dimensional  harmonic oscillator and the Milne--Pinney
equation with \mbox{$k=1$}, which  has the above-mentioned conditions. Its solutions are the integral curves of the
$t$-de\-pen\-dent vector f\/ield
\[X=v_x\frac{\partial}{\partial x}+v_y\frac{\partial}{\partial y}-\omega^2(t)x\frac{\partial}{\partial v_x}+\left(-\omega^2(t)y+\frac{1}
  {y^3}\right)\frac{\partial}{\partial v_y} ,
\]
which is a linear combination with time-dependent coef\/f\/icients,
 $X=X_2-\omega^2(t)X_1$, of
\[X_1=x\frac{\partial}{\partial v_x}+y\frac{\partial}{\partial v_y} ,\qquad X_2=v_x\frac{\partial}{\partial x}+v_y\frac{\partial}{\partial y}+\frac{1}
  {y^3}\frac{\partial}{\partial v_y} .
\]
This system closes on  a $\mathfrak{sl}(2,\mathbb{R})$ Lie algebra with $X_3$ given by
\[X_3=\frac 12\left(x\frac{\partial}{\partial
    x}+y\frac{\partial}{\partial y}-v_x\frac{\partial}{\partial v_x}
-v_y\frac{\partial}{\partial v_y}\right) .
\]

The generators of this Lie system
span a distribution of dimension two and there
is  no f\/irst-integral of the  motion for such subsystem. By adding the
other $\mathfrak{sl}(2,\mathbb{R})$ linear Lie system appearing in the Ermakov
system,
the harmonic oscillator with the same time-dependent angular frequency, as  the
rank of the space of the distribution in the 4-dimensional
space  is  three, there is an integral of motion.  The f\/irst-integral can
be obtained
from $X_1F=X_2F=0$. But $X_1F=0$ means that there exists a function $\bar
F:\mathbb{R}^3\to \mathbb{R}$ such that $F(x,y,v_x,v_y)=\bar F(x,y,\xi)$, with $\xi=xv_y-yv_x$,
and then $X_2F=0$ is written
\[
v_x\frac{\partial \bar F}{\partial x}+
v_y\frac{\partial \bar F}{\partial y}+\frac x{y^3}\frac{\partial \bar F}{\partial \xi}
\]
 and we obtain the associated system of characteristics
 \[\frac {x dy-y  dx}{\xi}=\frac {y^3 d\xi}{x}\Longrightarrow
  \frac{d(x/y)}{\xi}+\frac{y d\xi}{x}=0 ,
\]
from where the following f\/irst-integral is found \cite{Le67}:
\[
\psi(x,y,v_x,v_y)=\left(\frac{x}{y}\right)^2+\xi^2=\left(\frac{x}{y}\right)^2+
(xv_y-yv_x)^2 ,
\]
which is the well-known Ermakov--Lewis invariant \cite{RR79a,DL84,PGL91}.

\subsection{The Pinney equation revisited}

We mentioned before the possibility of obtaining solutions of a given system
from particular solutions of another related system. We next study a
 particular example which allows us to recover the results obtained
by  Pinney long time ago \cite{P50}.
 Consider the system of f\/irst-order dif\/ferential equations:
\begin{gather*}
\dot x = v_x,\\
\dot y = v_y,\\
\dot z = v_z,\\
\dot v_x = -\omega^2(t)x+\dfrac{k}{{x^3}},\\
\dot v_y = -\omega^2(t)y,\\
\dot v_z = -\omega^2(t)z,
\end{gather*}
which corresponds to the vector f\/ield
\[X=v_x\frac{\partial}{\partial x}+v_y\frac{\partial}{\partial y}+v_z\frac{\partial}{\partial z}+\frac{k}{x^3}
\frac{\partial}{\partial v_x}-\omega^2(t)\left(x\frac{\partial }{\partial v_x}+y\frac{\partial }{\partial v_y}+
z\frac{\partial}{\partial v_z}\right) .
\]
The $t$-dependent vector f\/ield $X$ can be expressed as $X=N_2-\omega^2(t)N_1$
where $N_1$ and $N_2$ are
\[
N_1=y\frac{\partial }{\partial v_y}+x\frac{\partial }{\partial v_x}+
z\frac{\partial}{\partial v_z},\qquad
 N_2=v_x\frac{\partial}{\partial x}+v_y\frac{\partial}{\partial y}+v_z\frac{\partial}{\partial z}+
\frac{1}{x^3}\frac{\partial}{\partial v_x},
\]
These vector f\/ields generate a 3-dimensional real Lie algebra with the
vector f\/ield $N_3$ given by
\[
 N_3=\frac 12\left(x\frac{\partial}{\partial x}+y\frac{\partial}{\partial y}
+z\frac{\partial}{\partial z}
-v_x\frac{\partial}{\partial
     v_x}
-v_y\frac{\partial}{\partial   v_y}
-v_z\frac{\partial}{\partial v_z}\right) .
\]
In fact, they generate a Lie algebra isomorphic to   $\mathfrak{sl}(2,\mathbb{R})$ because
 \[
[N_1,N_2]=2N_3, \qquad [N_3,N_1]=N_1, \qquad  [N_2,N_3]=N_2 .
\]

The dimension of the distribution generated by these fundamental vector f\/ields is three and the manifold of the Lie system is of dimension six, then there are three
 time-independent integrals of motion which turn out to be
the Ermakov invariant
 $I_1$ of the subsystem involving variables $x$ and $y$,
the Ermakov invariant $I_2$
of the subsystem involving variables~$x$ and~$z$, and the Wronskian $W$ of the subsystem involving variables $y$ and $z$.
They def\/ine a  foliation  with 3-dimensional leaves.
 This foliation  can be used for obtaining a superposition rule.

The Ermakov invariants read as
\[
I_1=\frac 12\left((yv_{x}-xv_y)^2+k\left(\frac {y}x\right)^2\right) ,\qquad
I_2=\frac 12\left((xv_{z}-zv_x)^2+k\left(\frac {z}x\right)^2\right) ,
\]
where $I_1$ and $I_2$ are non-negative constants and the Wronskian $W$ is:
\[
W=yv_{z}-zv_{y} .
\]
 We can obtain an explicit expression of $x$ in terms of $y$, $z$ and the
 three f\/irst integrals $I_1$, $I_2$, $W$:
\[
x=\frac {\sqrt 2}{\mid W\mid}\left(I_2y^2+I_1z^2\pm\sqrt{4I_1I_2-kW^2}\ yz\right)^{1/2} .
\]	

Remark that here $W$ is a constant f\/ixed by the two independent particular
solutions of the time-dependent harmonic oscillator $x_1(t)$ and $x_2(t)$ and
only $I_1$ and $I_2$
play the role of constants in this  superposition rule
for the Milne--Pinney equation. This is not a surprising fact
 because the Milne--Pinney equation is a second-order dif\/ferential equation. Note also that the values of~$I_1$ and~$I_2$ are  non-negative constants
but should be chosen such that  $x(0)$ be real.

This can be interpreted, as pointed out by Pinney \cite{P50},
 as saying that there is a superposition rule allowing
us to express  the general solution of the Milne--Pinney equation in terms of two
independent solutions of the corresponding harmonic oscillator with
the same  time-dependent angular frequency.

\subsection{A new superposition rule for the Pinney equation}

 A new  mixed superposition rule involving solutions
of a Riccati for obtaining the general solution of
Pinney equations is  obtained in this section
as a straightforward application of our development.

The $t$-dependent Riccati equation
\begin{gather}\label{ricceq}
\frac{dx}{dt}=b_0(t)+b_1(t)x+b_2(t)x^2 .
\end{gather}
 has been studied  in \cite{CarRam,CLR07b} from the perspective of the theory
of
Lie systems.  We follow here a~very similar approach. From the geometric
viewpoint
 the Riccati equation
(\ref{ricceq})  can be considered as a dif\/ferential equation
determining the integral curves of  the time-dependent vector
f\/ield
\begin{gather*}
 \Gamma=(b_0(t)+b_1(t)x+b_2(t)x^2)\frac{\partial}{\partial x}  ,%\label{vfRic}
\end{gather*}
which  is a linear combination with
time-dependent coef\/f\/icients of the three  vector f\/ields
\begin{gather*}
L_1 =\frac{\partial}{\partial x} ,  \qquad L_2 = -x^2
\frac{\partial}{\partial x}  ,  \qquad  L_3 =-x \frac{\partial}{\partial x},  %\label{sl2gen}
\end{gather*}
which close on a 3-dimensional real
Lie  algebra, with def\/ining relations{\samepage
\begin{gather} \label{conmutL}
[L_1,L_2] = 2L_3 ,
  \qquad [L_3,L_1] = L_1,\qquad [L_2, L_3] = L_2   ,
\end{gather}
therefore isomorphic to $\mathfrak{sl}(2,\mathbb{R})$,
because  the commutation relations
 (\ref{conmutL}) are the same as (\ref{CR}).}

The following
particular case of Riccati equation:
\begin{gather*}
\frac{dx}{dt}=-\omega^2(t)-x^2 ,
\end{gather*}
is the equation for the integral curves of the $t$-dependent vector f\/ield
$L=L_2-\omega^2(t)L_1$.
Consequently, we can apply the procedure of the Section~\ref{SR2} and consider
the following
system of dif\/ferential equations
\begin{gather*}
\dot x_1 =-\omega^2(t)-x_1^2,\\
\dot x_2 =-\omega^2(t)-x_2^2,\\
\dot x_3 =-\omega^2(t)-x_3^2,\\
\dot x =v,\\
\dot v =-\omega^2(t)x+\dfrac{k}{{x^3}}
\end{gather*}
described by a vector f\/ield in $\mathbb{R}^3\times {\rm T}\mathbb{R}_+$,
where $(x_1, x_2, x_3)\in \mathbb{R}^3$, $x\in \mathbb{R}_+$ and $(x,v)\in
T_x\mathbb{R}_+$. According to our general recipe, consider
the following vector f\/ields
\begin{gather*}
M_1  =\frac{\partial}{\partial x_1}+\frac{\partial}{\partial x_2}+\frac{\partial}{\partial x_3}+x\frac{\partial}{\partial v} , \qquad
M_2  = -x_1^2\frac{\partial}{\partial x_1}-x_2^2
\frac{\partial}{\partial x_2}-x_3^2
\frac{\partial}{\partial x_3}+v\frac{\partial}{\partial x}+\frac{k}{x^3}\frac{\partial}{\partial v},\\
M_3  =-x_1 \frac{\partial}{\partial x_1}-x_2 \frac{\partial}{\partial x_2}-x_3 \frac{\partial}{\partial x_3}+\frac{1}{2}\left(x\frac{\partial}{\partial x}-v\frac{\partial}{\partial v}\right) ,
\end{gather*}
that, by construction,  verify the same commutation relations as before, i.e.
\begin{gather*} %\label{conmutL2}
[M_1,M_2] = 2M_3 ,
  \qquad [M_3, M_1] = M_1,\qquad [M_3, M_2] = -M_2   ,
\end{gather*}
and the full system of  dif\/ferential equations can be understood as the
system of dif\/ferential equations for the determination of the integral curves
of the $t$-dependent vector f\/ield $M=M_2-\omega^2(t)M_1$. The dimension
of the distribution
associated
to this Lie system is three in almost any point and then there exist
 two integrals of motion. As $2 M_3=[M_1,M_2]$, it  is enough to f\/ind the
 simultaneously integrals of motion of $M_1$ and $M_2$, i.e.\ a function
$F:\mathbb{R}^5\rightarrow \mathbb{R}$ such that $M_1F=M_2F=0$.

Let us  f\/irst look for f\/irst-integrals independent of $x_3$, i.e.\ we  suppose
 that $F$ depends just on~$x_1$, $x_2$, $x$ and $v$.
Using the method of characteristics,  the condition $M_1F=0$ means that there is a function  $\bar F:\mathbb{R}^3\rightarrow \mathbb{R}$ such
that $F(x_1,x_2,x,v)=\bar F(I_1,I_2,I_3)$ with $I_1$, $I_2$ and $I_3$ given~by
\[
I_1=x_1-x_2 ,\qquad
I_2=x_2-v/x ,\qquad
I_3=x .
\]
Then, the condition $M_2\bar F=0$ reads in terms of the variables $I_1$, $I_2$, $I_3$ and $I_4\equiv v$ as
\begin{gather*}
I_4\left[-2\frac{I_1}{I_3}\frac{\partial\bar F}{\partial I_1}-2\frac{I_2}{I_3}\frac{\partial\bar F}{\partial I_3}+\frac{\partial\bar F}{\partial I_3}\right]+
\left[\left(-I_1^2-2I_1I_2\right)\frac{\partial \bar F}{\partial I_1}-\left(I_2^2+\frac{k}{I_3^4}\right)\frac{\partial\bar F}{\partial I_2}\right]=0.
\end{gather*}
Thus, the linear term in $I_4$ and the other one must vanish independently. The method of characteristics applied to the f\/irst term means that there exists a function $\widehat F:\mathbb{R}^2\rightarrow\mathbb{R}$ such that $\bar F(I_1,I_2,I_3)=\widehat F(K_1,K_2)$ where
\[
K_1=\frac{I_1}{I_2} ,\qquad
K_2=I_2I_3^2  .
\]
Finally, taking into account the last result in $M_2\hat F=0$ we obtain
\[
\left(-K_1^2-K_1+\frac{k K_1}{K_2^2}\right)\frac{\partial\widehat F}{\partial K_1}-\left(K_2+\frac{k}{K_2}\right)\frac{\partial\widehat F}{\partial K_2}=0,
\]
and by means of the method of characteristics
\[
\frac{dK_1}{dK_2}=\frac{K_1^2+K_1-\frac{k K_1}{K_2^2}}{K_2+\frac{k}{K_2}}.
\]
Finally, taking into account the last result we obtain the f\/irst-integral
\[
C_1=K_2+\frac{k+K_2^2}{K_1K_2} ,
\]
that in terms of the initial variables reads as
\[
C_1=\left(x_2-\frac{v}{x}\right)x^2+\frac{k+(x_2-\frac{v}{x})^2 x^4}{(x_1-x_2)x^2} .
\]
If we repeat the last procedure but under the assumption that the integral does not depend on~$x_2$ we obtain the following f\/irst-integral:
\begin{gather*}
C_2=\left(x_3-\frac{v}{x}\right)x^2+\frac{k+(x_3-\frac{v}{x})^2
  x^4}{(x_1-x_3)x^2}  .
\end{gather*}
It is a long but straightforward computation to check out that both are
integrals of
 $M_1$, $M_2$ and~$M_3$.
We can obtain now the general solution $x$ of (\ref{MP}) $x$ in terms of $x_1$, $x_2$, $x_3$, $C_1$, $C_2$ as
\begin{gather}
x=\sqrt{\frac{(C_1(x_1-x_2)-C_2(x_1-x_3))^2+k(x_2-x_3)^2}{(C_2-C_1)(x_2-x_3)(x_2-x_1)(x_1-x_3)}} ,\label{SupRul}
\end{gather}
where $C_1$ and $C_2$ are constants such that once $x_1(t)$, $x_2(t)$ and
$x_3(t)$
have been f\/ixed they make~$x(0)$ given by~(\ref{SupRul}) to be real.

Thus we have obtained a new mixed superposition rule which allows us to express
 the general solution of the Pinney equation in terms of three
solutions of Riccati equations, and of course two constants which determine
each particular solution.

\section{Conclusions and outlook}

In this paper we have reviewed the theory developed in \cite{CGM} in order to
deal with
a certain kind of systems second-order dif\/ferential equations
that  can be studied as Lie systems, in particular with the Milne--Pinney
equation
 and the Ermakov system. We have not only recovered  in this way some known
 results about these dif\/ferential equations, i.e. integrals of motion or
 superposition rules, but  we have also been able to show new ones as a
 superposition rule for three solutions of a Riccati equation to give us the
 general solution of the Milne--Pinney equation.

New applications of this formalism to
relate
a Milne--Pinney equation
with a frequency $\omega(t)$ with a $t$-dependent harmonic oscillator with
dif\/ferent frequency $\omega'(t)$ related with $\omega(t)$ by means of certain
relations
will be developed in forthcoming papers. In this way a new time-dependent
superposition rule is obtained and by means of it we can express
 the solution of a Milne--Pinney equation with frequency $\omega(t)$ in terms of solutions of
 a time-dependent harmonic oscillator with frequency $\omega'(t)$.

Another remarkable  point to be studied is the extension of this formalism
for studying some generalised Ermakov systems as those appearing  for instance
 in \cite{At99} and \cite{RSB96}.

\subsection*{Acknowledgements}
Partial f\/inancial support by research projects MTM2006-10531 and E24/1 (DGA)
 are acknow\-led\-ged. JdL also acknowledge
 a F.P.U.\ grant from  Ministerio de Educaci\'on y Ciencia and a~special grant from the Network of
Mechanics, Geometry and Control.

\pdfbookmark[1]{References}{ref}
\LastPageEnding

\end{document}